\documentclass[sigconf,screen]{acmart}

\usepackage{xspace}

\usepackage{xcolor}
\definecolor{dgrey}{gray}{0.2}

\definecolor{violet}{HTML}{6a51a3}

\usepackage{cleveref} 
\usepackage[inline]{enumitem} 

\usepackage{color, colortbl}  

\usepackage{adjustbox}
\usepackage{siunitx}
\usepackage{tabularx} 
\usepackage{enumitem} 
\usepackage{amsmath}  

\newcommand{\myparagraph}[1]{\paragraph*{\hspace*{-\parindent}\normalsize\bf#1}}

\usepackage{tcolorbox}

\title{Can We Hide Machines in the Crowd? Quantifying Equivalence in LLM-in-the-loop Annotation Tasks}

\author{Jiaman He}
\orcid{0009-0007-2817-7675}
\affiliation{%
\institution{RMIT University}
\city{Naarm/Melbourne}
\country{Australia}
}
\email{jiaman.he@student.rmit.edu.au}

\author{Zikang Leng}
\orcid{0000-0001-6789-4780}
\affiliation{%
\institution{Georgia Institute of Technology}
\city{Atlanta}
\country{USA}
}
\email{zleng7@gatech.edu}

\author{Dana McKay}
\orcid{0000-0001-7522-1842}
\affiliation{%
\institution{RMIT University}
\city{Naarm/Melbourne}
\country{Australia}
}
\email{dana.mckay@rmit.edu.au}

\author{Damiano Spina}
\orcid{0000-0001-9913-433X}
\affiliation{%
\institution{RMIT University}
\city{Naarm/Melbourne}
\country{Australia}
}
\email{damiano.spina@rmit.edu.au}

\author{Johanne R. Trippas}
\orcid{0000-0002-7801-0239}
\affiliation{%
\institution{RMIT University}
\city{Naarm/Melbourne}
\country{Australia}
}
\email{j.trippas@rmit.edu.au}

\renewcommand{\shortauthors}{Jiaman He, Zikang Leng, Dana McKay, Damiano Spina, and Johanne R. Trippas}

\copyrightyear{2025}
\acmYear{2025}
\setcopyright{cc}
\setcctype{by}
\acmConference[SIGIR-AP 2025]{Proceedings of the 2025 Annual International
ACM SIGIR Conference on Research and Development in Information Retrieval in
the Asia Pacific Region}{December 7--10, 2025}{Xi'an, China}
\acmBooktitle{Proceedings of the 2025 Annual International ACM SIGIR
Conference on Research and Development in Information Retrieval in the Asia
Pacific Region (SIGIR-AP 2025), December 7--10, 2025, Xi'an,
China}\acmDOI{10.1145/3767695.3769508}
\acmISBN{979-8-4007-2218-9/2025/12}

\begin{document}

\begin{abstract}
    
Many evaluations of large language models (LLMs) in text annotation focus primarily on the correctness of the output, typically comparing model-generated labels to human-annotated ``ground truth'' using standard performance metrics. In contrast, our study moves beyond effectiveness alone. We aim to explore how labeling decisions--by both humans and LLMs--can be statistically evaluated across individuals. Rather than treating LLMs purely as annotation systems, we approach LLMs as an alternative annotation mechanism that may be capable of mimicking the subjective judgments made by humans. To assess this, we develop a statistical evaluation method based on Krippendorff’s $\alpha$, paired bootstrapping, and the Two One-Sided t-Tests (TOST) equivalence test procedure. 
This evaluation method tests whether an LLM can blend into a group of human annotators without being distinguishable.  We apply this approach to two datasets, MovieLens 100K and PolitiFact, and find that the LLM is statistically indistinguishable from a human annotator in MovieLens 100K ($p = 0.004$), but not in PolitiFact ($p = 0.155$), highlighting task-dependent differences. 


\end{abstract}



\maketitle

\section{Introduction}
\label{sec:intro}

In Alan Turing's 1950 landmark 1950 paper, he proposed a criterion for machine intelligence: if a machine can engage in conversation such that a human evaluator cannot reliably distinguish it from another human. The machine can be said to exhibit intelligent behaviour~\cite{turing2009computing}. This formulation, now known as the Turing Test, shifts the focus from a machine's internal workings to its external behaviour.
In this work, we build on the spirit of the Turing Test. However, we focus on a more domain-specific setting: text annotation tasks. Instead of testing a machine through open-ended conversations, we ask a simpler question. Can an LLM act like a human annotator? Specifically, can its output be statistically indistinguishable from that of people in a multi-person annotation~task?

While LLMs have shown strong performance on many general-purpose classification tasks~\cite{nasution2024chatgpt, zhang-etal-2024-sentiment, hegselmann2023tabllm},  their role in subjective or domain-specific annotation settings remains unclear. In information retrieval contexts—such as document relevance assessment, intent classification, or stance detection—annotation often involves subtle, context-dependent judgments~\cite{faggioli2023perspectives, roitero2023relevance, kutlu2018crowd, schnabel2025multi}. These applications continue to rely on traditional classifiers trained on human-labeled data, particularly where interpretability, auditability, or fairness are required~\cite{lin2021truthfulqa, macavaney2023one}. If LLMs can produce labels that are indistinguishable from those generated by humans in such scenarios, they offer a path to reducing annotation costs while preserving human-like interpretive behavior. This makes it crucial to understand whether LLMs can accurately and effectively participate in the human annotation process used to train these models.

Text classification is widely used for natural language processing (NLP) tasks, such as news categorization, sentiment analysis, or subject labeling~\cite{dong2022survey, siino2025foundations}. It involves assigning labels to textual elements such as sentences, questions, paragraphs, or entire documents. Some of the classifications require human labeling to train and test the machine learning models. Annotations act as the ground truth against which models are tested, refined, and~advanced.

Annotation tasks often depend on subjective human judgment rather than a single, objective truth~\cite{magnossao2025effects}. For example, in relevance assessments, people may interpret the same content differently, and their judgments can change over time~\cite{faggioli2023perspectives}. So, the aim is not 
always to identify the ``correct'' label, but to understand how labeling decisions emerge across different individuals. LLMs are increasingly good at producing convincing outputs. However, it is 
still unclear whether their responses reflect human judgment, especially in the absence of evidence that their outputs are rooted in actual human experience~\cite{faggioli2023perspectives}.

Traditional human annotation can be constrained by cost and consistency~\cite{devillers2005challenges}. Recent research has compared the quality of annotations by LLM and humans for NLP applications~\cite{nasution2024chatgpt}. Also, a study investigated using LLMs to help or even replace human annotators on some tasks~\cite{alaofi2024llms},
usually by comparing their agreement with human results using measures like Krippendorff's $\alpha$ or Cohen's kappa. Most of these studies treat the LLM as a single, standalone system and check how well it matches a human-created ``ground truth''~\cite{macavaney2023one, alaofi2024llms, zendel2024enhancing, dietz2025llm,balog2025rankers}.

We introduce an evaluation method for LLMs based on group dynamics. Rather than evaluating a model in isolation, we assess whether it can substitute a human within an annotation group without significantly altering the group's behavior. An LLM judgment is deemed successful if the LLM's presence is statistically indistinguishable from a human's presence.

This work treats LLMs not just as tools for text classification, but as participants that can imitate the subjective and sometimes inconsistent judgments made by humans. We introduce a practical method based Krippendorff's $\alpha$, bootstrapping and TOST to test whether an LLM can blend into a group of human annotators without being identified. This approach requires only a small number of annotation items and functions as a domain-specific adaptation of the Turing Test. It supports early-stage evaluation on a small sample to determine the suitability of LLMs for large-scale annotation. We apply it to a real-world classification task and examine the results.
Our key contributions are:
\vspace{-0.05in}

\begin{itemize}
\item We propose an evaluation methodology that statistically tests whether an LLM can substitute for a human annotator in multi-annotator text classification tasks.
\item We demonstrate the application of our methodology on two datasets—MovieLens 100K and PolitiFact—showing that the LLM is statistically indistinguishable from human annotators in MovieLens 100K ($p = 0.004$) but not in PolitiFact ($p = 0.155$), revealing important task-dependent differences.  
\item We release a dataset containing LLM annotations alongside human annotations for a multi-annotator task. The dataset is publicly available at: \url{https://github.com/peanutH/LLM-evaluation}.
\end{itemize}

\section{Related Work}

\subsection{Text Classification}\label{subsec:text classification}

Text classification is a core task in NLP. It helps organize unstructured text from sources like messages, documents, and websites. To make sense of all this text, researchers and developers use models that automatically classify text into categories such as topic, sentiment, or author identity~\cite{siino2025foundations}.
%
In addition, labeled data is essential for building and testing models, with human annotations serving as the ``ground truth''~\cite{nasution2024chatgpt, siino2025foundations}. Manual labeling is often costly, slow, and inconsistent, making automation preferable. Models are typically trained on a subset of labeled data and evaluated by comparing their predictions to human annotations.

Recently, LLMs have shown impressive performance across many NLP tasks ~\cite{NEURIPS2020_1457c0d6}, raising the possibility of replacing traditional supervised models or even human annotators. However, this shift is far from complete~\cite{movva2024annotation}.
Many domain-specific real-world applications, such as medical coding, legal triage, or sentiment analysis, continue to rely on traditional classifiers trained on large datasets labeled by experts. This is because LLMs alone often lack the nuanced understanding required for these tasks. As a result, substantial human annotation remains essential for developing and validating reliable models. In areas where complex human judgment is critical, direct human involvement remains indispensable~\cite{christoforou2025crowdsourcing}.

\subsection{Human Annotation and Generative AI}
\label{subsec:crowdsourcedannotation}

Human annotation comes with challenges, most notably cost and consistency~\cite{devillers2005challenges}. Text classification models rely on large amounts of labeled data, and hiring workers to do all the labeling is not always feasible. That is why crowdsourcing has become a solution. Platforms that connect requesters with crowd workers make it possible to outsource labeling tasks, such as relevance judgments, sentiment tagging, and topic categorization, to non-experts~\cite{alonso2012using, kutlu2018crowd}. This approach has been crucial in building datasets needed to train machine learning models.

Crowdsourcing was once viewed as a flexible and empowering option for workers. However, it is often criticized as invisible, low-paid labor that supports modern AI behind the scenes.
Given these concerns, researchers have started asking: Can generative AI (GenAI), especially LLMs, step in and take over some of these annotation tasks? Some early findings suggest that LLMs tend to perform well on straightforward tasks, such as summarization or basic sentiment analysis~\cite{christoforou2025crowdsourcing}. But when the task requires more nuanced judgment—like interpreting sarcasm, ambiguity, or subtle context—human annotators still outperform the machines~\cite{nasution2024chatgpt}.

So, the real question is not just whether GenAI can get the ``right'' answer. It is whether its decisions reflect the kinds of judgments humans would make, especially in cases where there is no single correct label. One study~\cite{faggioli2023perspectives} examined the ability of LLMs to handle relevance judgments, a task where subjectivity plays a role. While LLMs showed some ability to mimic human responses, they were not consistent or nuanced enough to fully replace human workers.

That is why we are developing a new evaluation framework to better understand whether LLM-generated annotations can be mistaken for human ones, not just in terms of correctness but in how closely they match human reasoning and subjectivity.

\subsection{Existing Evaluation Method}
\label{subsec:existing evaluation method}

Traditional evaluation methods for generative AI in annotation tasks typically benchmark AI-generated outputs against human annotations using metrics such as accuracy, precision, Kendall’s $\tau$, or inter-annotator agreement scores like Cohen’s kappa and Krippendorff’s $\alpha$~\cite{faggioli2023perspectives, zendel2024enhancing, alaofi2024llms}. The approaches used in existing studies treat generative AI as a system and evaluate their output against the overall consensus of a crowd. In doing so, they prioritize alignment with collective human judgments, rather than examining how closely AI aligns with the characteristics of individual annotators.

Although these evaluations show whether generative models can produce generally accurate labels, they often miss how humans actually annotate. In real tasks—especially subjective ones like relevance, sentiment, or moderation—judgments vary with expertise, interpretation, or background \cite{mizzaro1997relevance}. Treating this diversity as a single “gold standard” can overstate model capability~\cite{faggioli2023perspectives}.

Correlation metrics like Kendall’s $\tau$~\cite{kendall1938new} are good for checking if LLM rankings match system-level outcomes. But they don’t show how well LLMs fit into the social side of annotation, where disagreement and variation are normal. Measures like Cohen’s $\kappa$~\cite{cohen1960coefficient} and Krippendorff’s $\alpha$~\cite{krippendorff2011computing} better capture consistency, but they still treat LLMs as outsiders compared to human annotators, rather than as active collaborators in the process.

This framing can lead to an inflated sense of how interchangeable LLMs are with humans, particularly in complex or cognitively demanding annotation settings~\cite{pavlick2019inherent}. As prior studies have noted, even as LLMs improve at mimicking human language and surface-level judgment, it remains a significant leap to assume their outputs are equivalent to human reasoning without verification. At present, there is no definitive evidence that LLM-generated judgments are grounded in human experience, intuition, or context.

This raises an important question: if an LLM’s annotation looks like a human's, does that mean it is truly the same, or are we missing differences in how judgments are made? In many tasks, there is no single ``correct'' answer; human judgments are often subjective, context-dependent, and inconsistent over time~\cite{bailey2008relevance, faggioli2023perspectives}.
To better reflect this, we propose a new approach: instead of checking if an LLM agrees with the crowd, we ask whether it can blend into the crowd—becoming statistically indistinguishable from human annotators, and that we call it a ``Statistical Turing Test''.

\subsection{Human Judgment}
\label{subsec:human judgement}

Before comparing LLMs to humans, it is important to first understand the nature of human judgment. Human judgment is commonly modeled as a cognitive process that aligns well with linear models of cue integration~\cite{hammond1955probabilistic, hoffman1960paramorphic}. In such models, people make decisions based on a set of cues, each weighted differently depending on its perceived importance.

\citet{brehmer1980cognitive} noted that linear models tend to fit human judgments quite well. Even when nonlinear or configural components are present, they typically account for only a small portion of the variance, and their generalizability across tasks is uncertain. Additionally, human judgments are often inconsistent, with the level of consistency varying according to the predictability of the task. There are also substantial inter-individual differences in how people weigh signals, even among individuals with considerable experience on the task.

This inconsistency can be attributed to the variability in cue weights applied across different tasks. Prior research has shown that judgment consistency tends to decrease as the number of cues increases~\cite{einhorn1971use}. In contrast, LLMs often display greater consistency in annotation tasks~\cite{faggioli2023perspectives}, likely due to more stable internal representations of cues and weights.
To evaluate whether the LLM's cue integration falls within an acceptable threshold of variability, we use inter-annotator agreement (IAA) as a proxy for measuring consistency in annotation judgments.

\subsection{Inter-Annotator Agreement (IAA)}
\label{subsec:inter annotator agreement}

Researchers who rely on hand-labeled data, where items are manually labeled with categories for empirical analysis or model development, must demonstrate that the labeling process is reliable~\cite{artstein2008inter}. A fundamental assumption in annotation methodology is that the data are considered reliable when multiple annotators agree on the labels assigned, to a degree appropriate for the objectives of the study~\cite{krippendorff2018content, craggs2004two}. 

Consistent agreement among annotators suggests that they share a common understanding of the annotation guidelines, and thus can be expected to apply those guidelines consistently.
IAA is a metric used to quantify this consistency. In multi-people annotation settings, annotators may have varied backgrounds and limited domain expertise. IAA helps determine whether labels are trustworthy and whether a task is clearly defined or inherently subjective. High agreement indicates clear instructions and straightforward data, while low agreement may reveal task ambiguity, multiple valid interpretations, or inconsistent annotator behavior.

Beyond assessing label quality, IAA also serves as a diagnostic tool for identifying issues in the annotation process. By examining patterns of agreement and disagreement, researchers can uncover sources of ambiguity, identify annotator bias, and refine the guidelines. One of the most used IAA measure is Krippendorff's $\alpha$. 

\subsection{Krippendorff’s \texorpdfstring{$\alpha$}{alpha}}
\label{subsec:krippendorffAlpha}
Krippendorff’s $\alpha$ is a robust and widely-used reliability coefficient for measuring inter-annotator agreement, particularly when annotations are incomplete, involve more than two coders, or span different levels of measurement (nominal, ordinal, interval, etc.)~\cite{krippendorff2011computing}. Unlike simpler metrics such as Cohen’s kappa, which assume fixed pairwise comparisons, Krippendorff’s $\alpha$ can accommodate complex and realistic annotation setups—including crowdsourced data with missing entries or unequal contributions from annotators.

Krippendorff’s $\alpha$ quantifies the extent to which observed disagreement differs from what would be expected by chance, with values ranging from 1 (perfect agreement) to 0 (chance-level agreement) and negative values indicating systematic disagreement. Importantly, Krippendorff’s $\alpha$ is sensitive not only to consistency but also to the nature of the scale being used, making it well-suited for subjective or ambiguous tasks where subtle distinctions matter.

In our evaluation framework, we employ Krippendorff’s $\alpha$ to assess whether annotations produced by an LLM achieve a comparable level of agreement with human annotators as humans achieve with one another. Rather than simply comparing the LLM to a gold standard, we integrate it into the annotator pool and compute Krippendorff’s $\alpha$ across the mixed group.

\section{Methodology}
\label{methodology}

Building on the idea discussed in Section~\ref{subsec:inter annotator agreement}, consistent labeling by multiple human annotators suggests that they share a common understanding of the annotation guidelines and apply them reliably~\cite{artstein2008inter}. Inspired by the logic of the Turing Test~\cite{turing2009computing}, we propose a methodology that uses inter-annotator agreement to evaluate whether an LLM can function as an individual annotator. That is, if it can serve as a substitute for a human in the annotation process.

\begin{figure}[t]
    \centering
        \includegraphics[width=\linewidth]{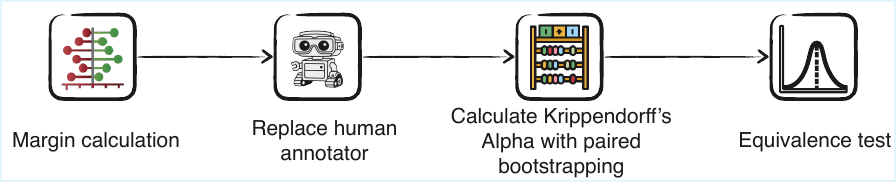}
    \caption{Our evaluation methodology workflow
    }
    \label{fig:evaluation_workflow}
\end{figure}

In this section, we outline the methodology used to evaluate whether an LLM can effectively substitute for a human annotator by examining changes in inter-annotator agreement, the methodology workflow is shown in~\autoref{fig:evaluation_workflow}. We first introduce the rationale behind this evaluation (Section~\ref{sec:Motivation: Substituting Human Annotators}). We then detail the protocol for substituting a human annotator with an LLM (Section~\ref{sec:Protocol: Replacing Annotators with an LLM}), followed by the approach for measuring how agreement levels vary due to these substitutions (Section~\ref{sec:Measuring Agreement Across Substitutions}). Next, we describe our statistical procedures for estimating variability in agreement scores using a paired bootstrap method (Section~\ref{sec:paired boostrap}) and establishing equivalence through Two One-Sided Tests (TOST) (Section~\ref{sec:equivalence testing with TOST}). Finally, we discuss practical considerations for determining appropriate sample sizes and annotator group sizes to ensure robust and reliable results (Section~\ref{sec:Sample size and Annotator Group Size}).

\subsection{LLM Substitution Protocol}
\label{sec:LLM substitution protocol}
\subsubsection{Motivation: Substituting Human Annotators}
\label{sec:Motivation: Substituting Human Annotators}
Krippendorff's $\alpha$ measures the extent of agreement among annotators using the formula:
\[
\alpha = 1 - \frac{D_o}{D_e}
\]
where $D_o$ is the observed disagreement and $D_e$ is the expected disagreement by chance.

To understand how $\alpha$ behaves under substitution, consider a group of three annotators-A, B, and C—whose annotations yield an agreement score $\alpha_1$.  Now imagine replacing annotator A with a new annotator E and computing a new $\alpha$ value, $\alpha_2$. If $\alpha_1 \approx \alpha_2$, this suggests that annotator E exhibits a similar consistency pattern to annotator A for the same task. Repeating this comparison with different annotators and observing small differences (i.e., $|\alpha_1 - \alpha_2|$ within a tolerable range) may indicate that the consistency patterns among the annotators are comparable.

This idea motivates our approach: if an LLM can replace a human annotator without significantly altering the inter-annotator agreement, it may be acting as a reasonable substitute.

\begin{figure}[t]
    \centering
    \includegraphics[width=\linewidth]{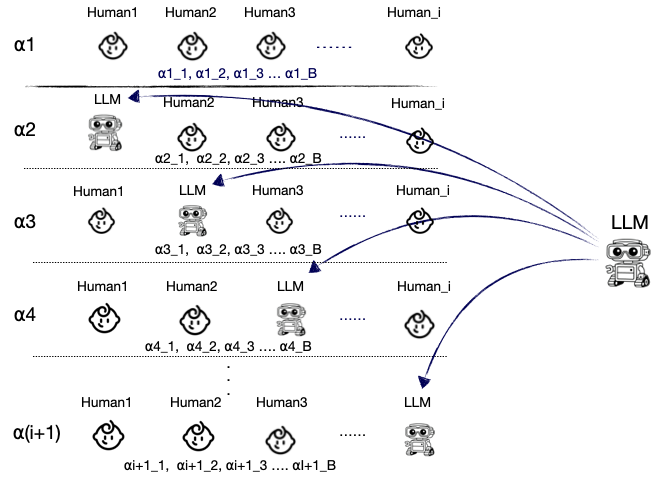}
    \caption{LLM substitution
    }
    \label{fig:LLM Substitution}
\end{figure}


\subsubsection{Protocol: Replacing Annotators with an LLM}
\label{sec:Protocol: Replacing Annotators with an LLM}

To formalize this idea, we consider a group of $i$ human annotators who have independently labeled a shared set of $n$ items. We then simulate the substitution process by replacing one human annotator at a time with the LLM. This results in $i$ modified annotation groups—each with $i-1$ humans and one LLM.

In each iteration, we remove the annotations from one human (e.g., annotator A) and replace them with labels generated by the LLM for the same items. The LLM effectively stands in for the removed annotator. This process is repeated for all $i$ human annotators. The substitution progression is illustrated in Figure~\ref{fig:LLM Substitution}.

In cases where the original annotator did not label all items, we only substitute the entries that exist—i.e., the LLM only replaces the ratings for items the original annotator labeled. Items left blank by the original annotator remain blank.

\subsubsection{Measuring Agreement Across Substitutions}
\label{sec:Measuring Agreement Across Substitutions}

In complex annotation tasks involving comprehension, reasoning, or subtle interpretation, perfect agreement is unlikely—even among humans.  We therefore don’t expect the LLM to match human annotations exactly. Instead, we evaluate whether the LLM can match the overall consistency of human annotators. 


We begin by computing Krippendorff’s $\alpha$ for the original group of $i$ human annotators. We denote this baseline agreement as:

\[
\alpha_1 = \text{Krippendorff's $\alpha$ for human group}
\]

Next, we compute $\alpha$ for each of the $i$ modified groups where one human annotator is replaced by the LLM. These are denoted as:

\[
\alpha_j = \text{Krippendorff’s $\alpha$ for substitution group} j = 2, \dots, i+1
\]

By comparing the LLM-substituted scores $\alpha_j$ with the original human score $\alpha_1$, we can observe how agreement changes when the LLM replaces a human annotator in the annotation group.

\subsubsection{Paired Bootstrap for Variability Estimation}
\label{sec:paired boostrap}

To assess the variability of agreement scores and ensure that observed differences are statistically meaningful, we apply the paired bootstrap method inspired by~\citet{krippendorff2011computing} 

Prior work has shown that analyzing a subset comprising just 40\% of the full dataset can  provide a reliable estimates of inter-rater agreement~\cite{armstrong2020accuracy}. Following this insight, we apply bootstrap sampling~\cite{tibshirani1993introduction} to resample the annotation data. 

Specifically, we perfrom $B$ bootstrap iterations (eg., $B=300$), where in each iteration we sample $N$ items with repalcement from the full set of $n$ annotated items. For each sample, we compute the Krippendorff's $\alpha$.

 Let $\alpha_{1}^{(1)}, \alpha_{1}^{(2)}, \dots, \alpha_{1}^{(B)}$ represent the $\alpha$ values computed for the original human group across the $B$ bootstrap samples. 

\begin{equation}
\alpha_1 = \left\{\alpha_{1}^{(1)}, \alpha_{1}^{(2)}, \dots, \alpha_{1}^{(B)}\right\}
\end{equation}

For each LLM-substituted group $j=2,\dots,i+1$,  we similarly compute a distribution of alpha values:

\begin{equation}
\alpha_j = \left\{\alpha_{j}^{(1)}, \alpha_{j}^{(2)}, \dots, \alpha_{j}^{(B)}\right\}
\end{equation}


To ensure fair comparison, we use the same bootstrap samples (i.e., same sampled items) across all groups in each iteration. This paired bootstrap procedure allows us to compare the variability in agreement across human and LLM-substituted groups under consistent sampling conditions.

\subsection{Equivalence Testing with TOST}
\label{sec:equivalence testing with TOST}

We then summarize the results by computing the mean agreement score for the original human–human annotations:
\[
x_2 = \frac{1}{B} \sum_{f=1}^{B} \alpha_{1}^{(f)}
\]

and the average agreement score across all LLM–human replacement cases:
\[
x_1 = \frac{1}{i \cdot B} \sum_{j=2}^{i+1} \sum_{f=1}^{B} \alpha_{j}^{(f)}
\] 

Here, $x_1$ represents the overall mean of $\alpha_2, \alpha_3, \dots, \alpha_{i+1}$, capturing average agreement when one human is replaced by the LLM. $x_2$ represents the baseline agreement among all-human groups.

These means serve as the basis for an equivalence test using the Two One-Sided t-Tests (TOST) procedure \cite{sakai2025my,snapinn2000noninferiority,rogers1993using}. The goal is to determine whether the difference $|x_1 - x_2|$ falls within a pre-defined equivalence margin $\Delta_{\text{equiv}}$, which indicates a practically negligible difference in reliability.

We compute the two TOST statistics as follows:

\[
t_1 = \frac{(x_1 - x_2 - \Delta_{\text{equiv}})}{s \sqrt{\frac{1}{n_1} + \frac{1}{n_2}}}
\quad \text{and} \quad
t_2 = \frac{(x_1 - x_2 + \Delta_{\text{equiv}})}{s \sqrt{\frac{1}{n_1} + \frac{1}{n_2}}}
\]

where $s$ is the pooled standard deviation of the two sets of$\alpha$scores, and $B_1$, $B_2$ are the sample sizes of the two groups. 

The null hypothesis is that the difference exceeds the equivalence margin:

\[
H_0: |x_1 - x_2| > \Delta_{\text{equiv}}
\]

We reject $H_0$ if both $t_1$ and $t_2$ fall within the critical region for their respective one-sided tests, thereby concluding that the observed difference is within an acceptable range of equivalence.

\subsubsection{Equivalence Margin Definition}
\label{subsec:equivalence margin definition}

The equivalence margin $\Delta_{\text{equiv}}$ is a threshold below which the difference is considered negligible. We estimate this margin empirically based on natural variability among human annotators:

\[
\Delta_{\text{equiv}} = (\alpha_a - \alpha_b) \cdot \text{fraction}
\]

where:
\begin{itemize}
    \item $\alpha_a$ is Krippendorff’s $\alpha$ calculated from a group of human annotators (e.g., annotators 1–3),
    \item $\alpha_b$ is Krippendorff’s $\alpha$ from another independent group (e.g., annotators 4–6),
    \item Both groups annotate the same items using the same guidelines,
    \item $\alpha_a - \alpha_b$ reflects typical human-to-human variability,
    \item \texttt{fraction} is a scaling factor (e.g., 0.5 or 0.8) that controls how strict the equivalence test is — smaller values require the LLM to match humans more closely.
\end{itemize}

By grounding the margin in actual human variability, this approach makes the equivalence test realistic and interpretable. The margin reflects what is already tolerated in human-human comparisons, rather than relying on arbitrary cut-offs.

\subsection{Sample Size and Annotator Group Size}
\label{sec:Sample size and Annotator Group Size}

\subsubsection{Dataset Size}
\label{sec:Dataset Size}

To determine an appropriate minimum bootstrap sample size for calculating Krippendorff’s alpha, we follow Bloch and Kraemer’s formula~\cite{krippendorff2018content}, which takes into account the desired minimum agreement level $\alpha_{\min}$, a confidence level $z$, and the probability of observing agreement by chance $p_c$:

\[
N = z^2 \left( \frac{(1 + \alpha_{\min})(3 - \alpha_{\min})}{4(1 - \alpha_{\min}) p_c (1 - p_c)} \right)
\]

Following the paired bootstrap procedure outlined in Section~\ref{sec:paired boostrap}, we randomly sample N items from the dataset in each bootstrap iteration. 
Using \( z = 0.95 \),  \( \alpha_{\min} = 0.8\), and $p_c = 0.17$,  we calculate the minimum required size of each bootstrap sample to be \( N = 32 \). As noted in Section~\ref{sec:paired boostrap}, a bootstrap sample comprising 40\% of the full dataset is sufficient to yield a reliable estimate of interrater agreement. This implies that the full dataset should contain at least \( n = 2.5N = 80 \) items.

\subsubsection{Annotator Group Size}
\label{sec:Dataset Size}

We analyze how Krippendorff’s $\alpha$ changes when one human annotator is replaced by an LLM. Let $i$ be the number of annotators, $n$ the number of items, and $\ell_{ak}$ the label from annotator $a$ on item $k$. As mentioned in Section~\ref{subsec:krippendorffAlpha}, Krippendorff’s $\alpha$ is defined as:
\[
\alpha = 1 - \frac{D_o}{D_e},
\]
where $D_o$ is the average observed pairwise disagreement and $D_e$ is the expected disagreement under random labeling based on marginal label distributions.

\myparagraph{Change in observed disagreement.}
When coder $r$ is replaced by an LLM that assigns labels $L_k$, the change in $D_o$ becomes:
\[
\Delta D_o = \frac{2}{n\,i} \sum_{k=1}^n \left( \bar d_{\text{LLM},k} - \bar d_{r,k} \right),
\]
where $\bar d_{\text{LLM},k}$ and $\bar d_{r,k}$ denote the average disagreement between the LLM (or original coder $r$) and all other annotators on item $k$.

\myparagraph{Change in expected disagreement.}
Substituting in the LLM slightly alters the marginal distribution of labels, shifting $p_c \mapsto p'_c = p_c + \Delta p_c$. The first-order change in $D_e$ is then:
\[
\Delta D_e \approx -2 \sum_c p_c \Delta p_c.
\]

\myparagraph{Total change in $\alpha$.}
Using a first-order Taylor expansion of $\alpha$, we obtain:
\[
\Delta \alpha \approx -\frac{1}{D_e} \Delta D_o + \frac{D_o}{D_e^2} \Delta D_e.
\]
Substituting the expressions above yields:
\[
\Delta \alpha
\approx
-\,\frac{2}{n\,i\,D_e}
\sum_{k=1}^{n}\bigl(\bar d_{\text{LLM},k}-\bar d_{r,k}\bigr)
\;-\;
\frac{2\,D_o}{D_e^2}
\sum_{c}p_c\,\Delta p_c.
\]

The derivation\footnote{Full derivation can be found in \url{https://github.com/peanutH/LLM-evaluation}} shows that $\Delta\alpha$, resulting from substituting one human annotator with an LLM, is inversely proportional to the number of annotators $i$ (i.e., $\Delta\alpha \propto \frac{1}{i}$). This implies that as the number of annotators increases, the impact of a single substitution on $\alpha$ becomes smaller. To select an appropriate group size, we identify the elbow point on the curve of $\Delta\alpha$ versus the number of annotators, where the rate of change drops sharply. We apply the L-method~\cite{Salvador2004lmethod} to detect this point by fitting two lines to the curve—one before and one after each candidate split—and choosing the split that minimizes the total fitting error. This method captures the transition from rapid to gradual change.

\section{Experimental Evaluation}
\label{sec:experimental methodology}

\begin{figure}[t]
    \centering
        \includegraphics[width=\linewidth]{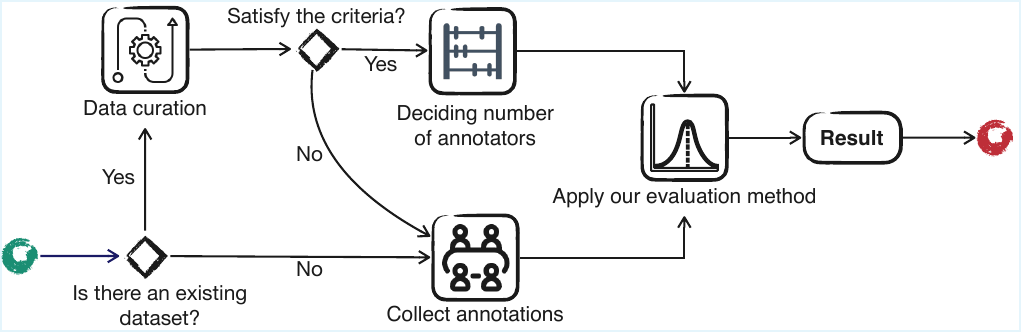}
    \caption{Guideline for applying our methodology
    }
    \label{fig:experiment_workflow}
\end{figure}

To validate our evaluation method, we designed an experimental workflow, illustrated in~\autoref{fig:experiment_workflow}. Since our experiments are conducted on existing datasets, we follow the corresponding branch of the workflow.

\subsection{Dataset}
\label{subsec:dataset}

\begin{table}[!t]
\centering
\caption{Datasets considered for evaluation.}
\resizebox{0.47\textwidth}{!}{%
\begin{tabular}{@{} l l r @{}}
\toprule
\textbf{Dataset} & \textbf{Domain} & \textbf{\# Items} \\
\midrule
MovieLens\,100K~\cite{harper2015movielens} & Movie ratings & 1,682 \\
WebCrowd25k~\cite{Kutlu2018webcrowd} & IR relevance & $\approx$4,500 \\
TREC-8 Re-assessments~\cite{roitero2021trec} & IR relevance & 4,269 \\
Familiarity–QuerySpec~\cite{he2025characterising} & Query specificity classification & 83 \\
MBIC~\cite{spinde2021mbicmediabias} & Media bias & $\approx$2,600 \\
CoQA~\cite{reddy2019coqa} & Conversational QA & 8,000 \\
POPQUORN~\cite{pei2023popquorn} & Offensive–QA & $\approx$5,500 \\
D3CODE~\cite{mostafazadeh2024d3code} & Cross-cultural offensiveness detection & $\approx$4,500 \\
CrowdsourcingTruthfulness--PolitiFact~\cite{roitero2020can} & Misinformation -- Veracity classification & 120 \\
HateXplain~\cite{mathew2021hatexplain} & Hate speech detection & $\approx$20\,000 \\
CODA-19~\cite{huang2020coda} & COVID-19 abstract section labeling & 10,966 \\
\bottomrule
\end{tabular}%
}
\label{tab:dataset-suitability}
\end{table}

We compiled a list of publicly available datasets that provide multiple annotations per item along with identifiable annotator IDs, enabling us to trace which individual labeled each item and to perform annotator-level substitutions. These datasets are summarized in~\autoref{tab:dataset-suitability}.
To fit our evaluation experiments, we filtered the datasets using a set of selection criteria defined by our evaluation methodology. The code for this filtering process is available.\footnote{Our filtering code can be found in~\url{https://github.com/peanutH/LLM-evaluation}} Specifically, we required that the dataset contain two disjoint groups of annotators, each of size $i$, such that there are at least 80 items, with each item annotated by at least two annotators from each group. After data filtering, two datasets satisfy our criteria: MovieLens 100K~\cite{harper2015movielens} and PolitiFact~\cite{roitero2020can}.

\label{subsec:dataset selection}

The MovieLens 100K dataset, collected by the GroupLens Research Project at the University of Minnesota~\cite{harper2015movielens}, contains 100,000 movie ratings (on a 1–5 scale) from 943 users across 1,682 movies. Each user rated at least 20 movies. The dataset also includes basic demographic information about the users, such as age, gender, occupation, and ZIP code.

The original PolitiFact dataset collected by~\citet{wang2017liar} contains 12,000 statements produced by U.S. politicians, each statement is labeled by an expert judge on a six-level scale for the statement's truthfulness. \citet{roitero2020can} selected 120 statements, 20 for each truth level, related to COVID-19 from the original PolitiFact dataset. Then, workers were recruited from Amazon Mechanical Turk to annotate each statement. Overall, each statement was annotated by 10 workers over three different scales: three-level, six-level, one-hundred level. In this work, we use the three-level dataset.  

To obtain the LLM annotations, we used GPT-4o mini to annotate the same set of items selected for this experiment, as described in Section~\ref{subsec:data selection}. We designed the prompt, see~\Cref{fig:prompt-template}, to follow the same annotation guidelines provided to human annotators. 

\begin{figure}[h]
\begin{tcolorbox}[colback=white,colframe=black,boxrule=1pt,arc=1mm]
\small
--SYSTEM--\newline
You are an average movie watcher. Rate each movie 
from 1 to 5 based on how much you liked it overall.
Consider the story, acting, and overall enjoyment.\newline

--USER--\newline
Respond with a list of ratings, one for each movie,
in the same order as presented. Only include the 
numeric ratings and nothing else.\newline

Here are the movies for you to rate:\newline
<LIST\_OF\_MOVIES>
\end{tcolorbox}
\captionsetup{skip=6pt} 
    \caption{LLM prompt for generating annotations for the MovieLens dataset. Further information is available on the paper's GitHub.}
    \vspace{-0.18in}
    \label{fig:prompt-template}
\end{figure}

\begin{figure}[t]
    \centering
        \includegraphics[width=0.95\linewidth]{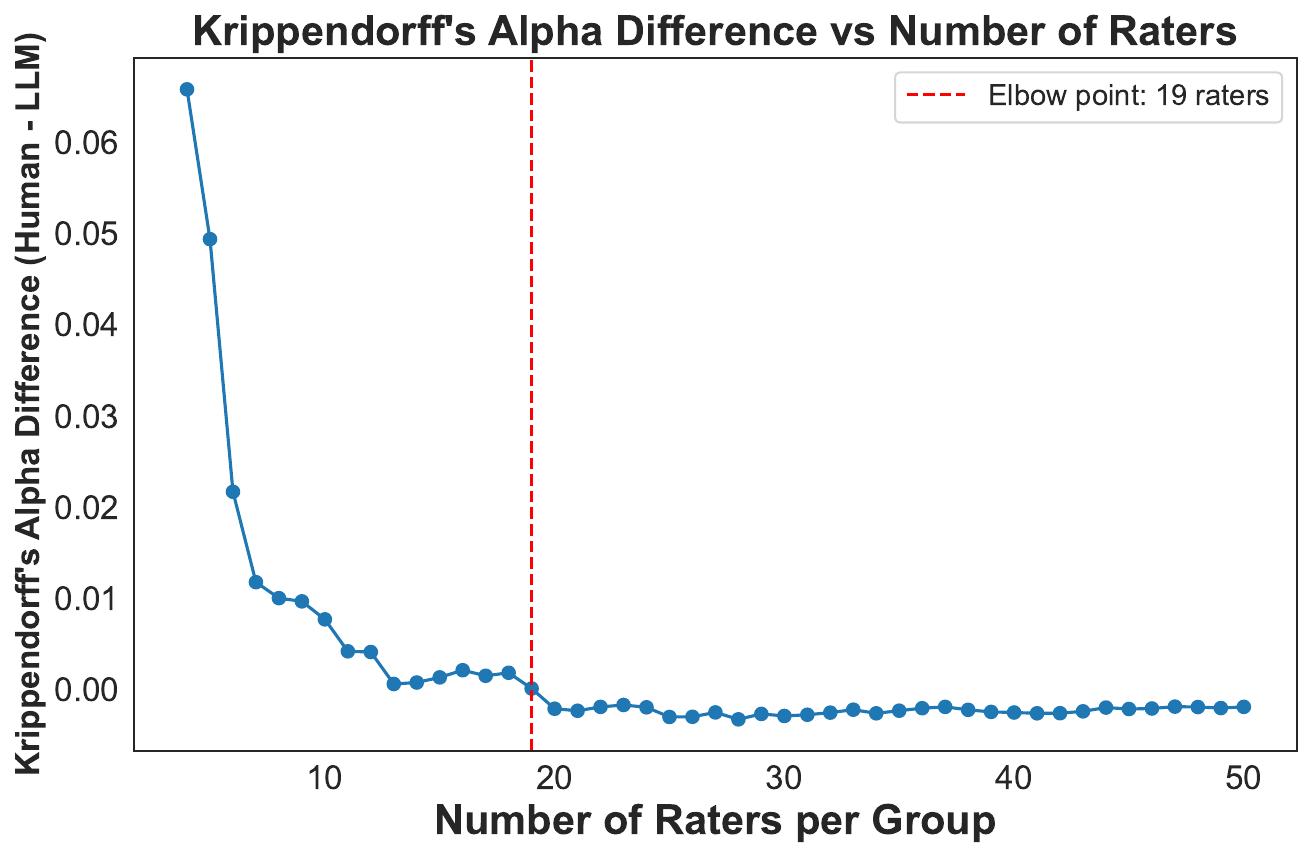}
    \caption{$p$-values obtained for the MovieLens 100K dataset for various value of $i$, the number of annotators in the group.  
    }
    \label{fig:movie_len_alpha_vs_num_raters}
\end{figure}


\subsection{Data Selection}
\label{subsec:data selection}

\begin{figure}[t]
    \centering
        \includegraphics[width=\linewidth]{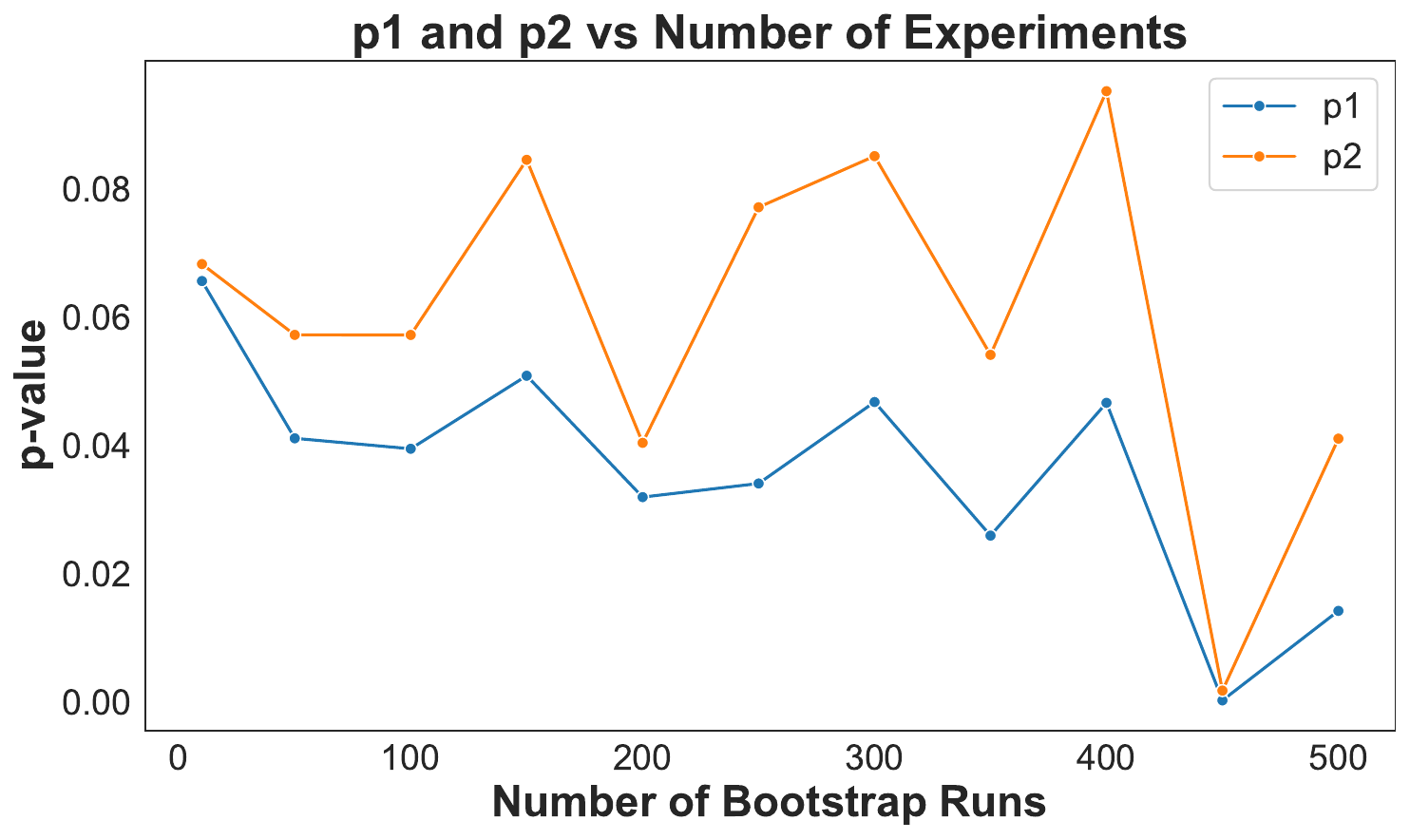}
    \caption{$p$-values obtained for the MovieLens 100K dataset for various value of $B$, the number of bootstrap iterations.  
    }
    \label{fig:movielen_p_evlauation}
\end{figure}


\myparagraph{MovieLens 100k}
For evaluation, we selected 38 coders split into two groups of 19 coders each. We used the elbow point finding method detailed in Section~\ref{sec:Sample size and Annotator Group Size} to determine the optimal number of annotators in a group. As shown in \autoref{fig:movie_len_alpha_vs_num_raters}, this point was at 19 annotators per group. 
We divided the coders into two groups:

\begin{itemize}
\item \textbf{Group A:} Human\textsubscript{1}, to Human\textsubscript{19}
\item \textbf{Group B:} Human\textsubscript{20} to Human\textsubscript{38}
\end{itemize}

\myparagraph{CrowdsourcingTruthfulness--PolitiFact}
We selected 86 coders for the evaluation—the minimum number that could be filtered from the dataset while still satisfying our evaluation criteria. The coders were divided evenly into two groups:

\begin{itemize}
\item \textbf{Group A:} Human\textsubscript{1} to Human\textsubscript{43}
\item \textbf{Group B:} Human\textsubscript{44} to Human\textsubscript{86}
\end{itemize}

This grouping enables the calculation of the equivalence margin (see Section~\ref{subsec:equivalence margin definition}). To satisfy the requirements for applying Krippendorff’s $\alpha$ that each item must be annotated by at least two coders within a group, and each coder must annotate at least one item~\cite{krippendorff2018content}, we filtered a subset of 100 items from the MovieLens 100k dataset and the PolitiFact dataset, each annotated by the selected coders. This meets the minimal number of items discussed in Section~\ref{sec:Sample size and Annotator Group Size}.

\begin{figure}[t]
    \centering
        \includegraphics[width=\linewidth]{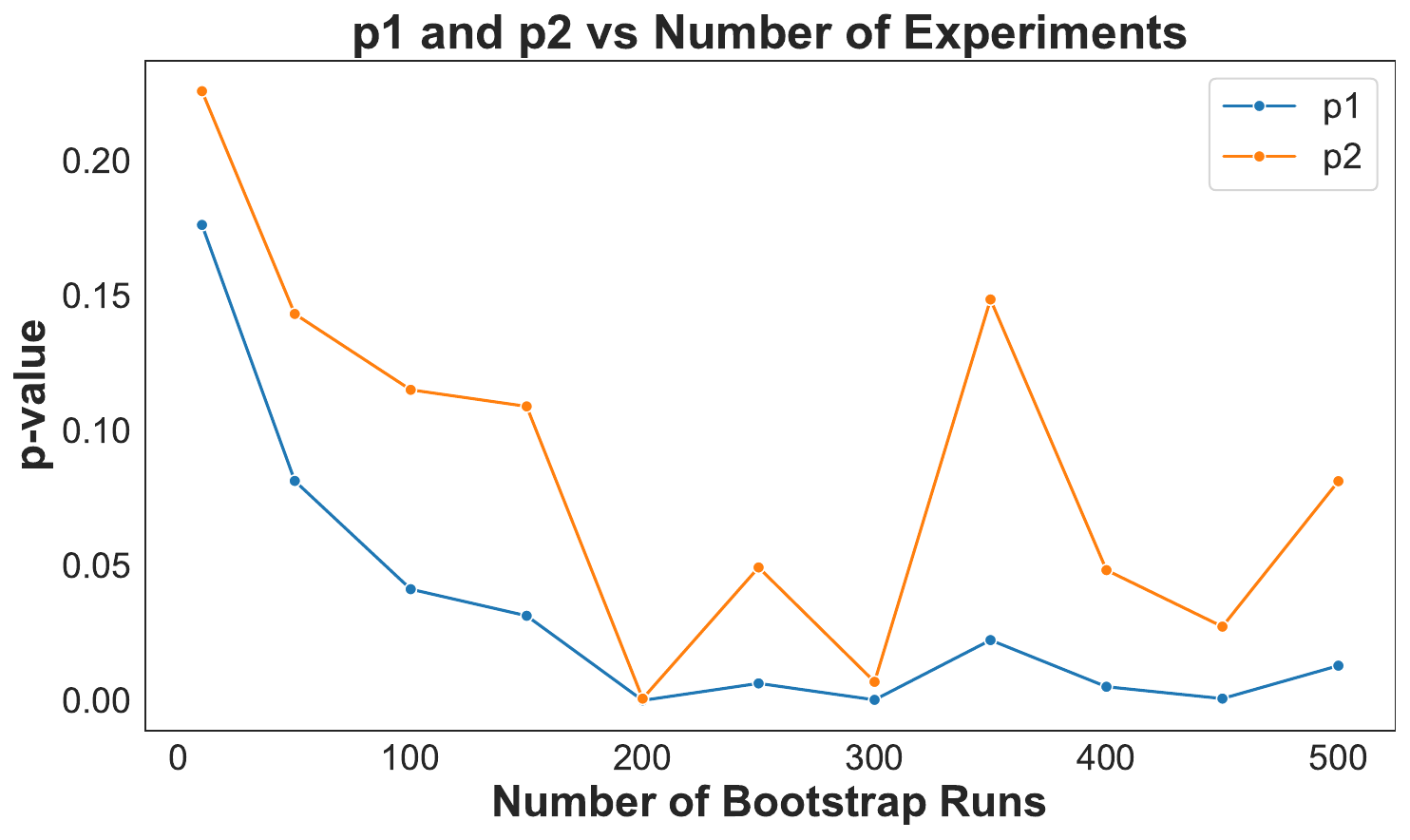}
    \caption{$p$-values obtained for the PolitiFact dataset for various value of $B$, the number of bootstrap iterations.  
    }
    \label{fig:Politifact_p_evaluation}
\end{figure}


\subsection{LLM Substitution Evaluation}
\label{subsec:llm substitution evaluation}

We evaluated whether LLMs could substitute for human annotators using the protocol introduced in Section~\ref{sec:LLM substitution protocol}. In each trial, we replaced one human coder at a time from Group A with an LLM, and computed Krippendorff’s $\alpha$ for the resulting group.

To assess variability and statistical significance, we applied the paired bootstrap procedure described in Section~\ref{sec:paired boostrap}. We tested multiple bootstrap sample sizes, with $B \in \{50, 100, \ldots, 500\}$. For each value of $B$, we repeated the substitution experiment \textbf{10 times}. Each repetition involved generating $B$ bootstrap samples of 40 items per iteration, computing agreement scores, and performing the Two One-Sided t-Test (TOST) as described in Section~\ref{sec:equivalence testing with TOST}.
For each value of $B$, we report the \textit{mean} and \textit{standard deviation} over the 10 trials for the two p-values from the TOST procedure ($p_1$ and $p_2$).

We also conducted a control experiment in which human annotations were replaced with randomly generated labels, rather than LLM-generated ones. This allowed us to assess how random substitution affects inter-rater agreement and to compare its impact against that of LLM substitution.

\subsection{Result}
\label{sec:result}

\begin{table*}[h]
    \centering
    \caption{TOST result ($\mu\pm\sigma$) for two datasets experiment for $B$ = 300}
    \adjustbox{max width=0.9\textwidth}{
    \begin{tabular}{lcccccccc}
        \toprule
        {\textbf{Dataset}} & {\textbf{Margin}} & {\textbf{Human $\alpha$}} &  {\textbf{LLM $\alpha$}} & {\textbf{Random $\alpha$}}  & {\textbf{LLM $p_1$}} & {\textbf{LLM $p_2$}} & {\textbf{Random $p_1$}} & {\textbf{Random $p_2$}} \\
        \midrule
         MovieLens 100K & 0.025 $\pm$ 0.014 & 0.199 $\pm$ 0.001 & 0.199 $\pm$ 0.001 & 0.164 $\pm$ 0.002 & 0.002 & 0.004 & 0.505  & <0.001 \\
         CrowdsourcingTruthfulness--PolitiFact & 0.034 $\pm$ 0.030 & 0.098 $\pm$ 0.004 & 0.100 $\pm$ 0.004 & 0.090 $\pm$ 0.004 & 0.047  & 0.155 & 0.092 & <0.001 \\
        \bottomrule
    \end{tabular}
    }
    \label{tab:p value result}
\end{table*}

\begin{figure}[t]
    \centering
        \includegraphics[width=\linewidth]{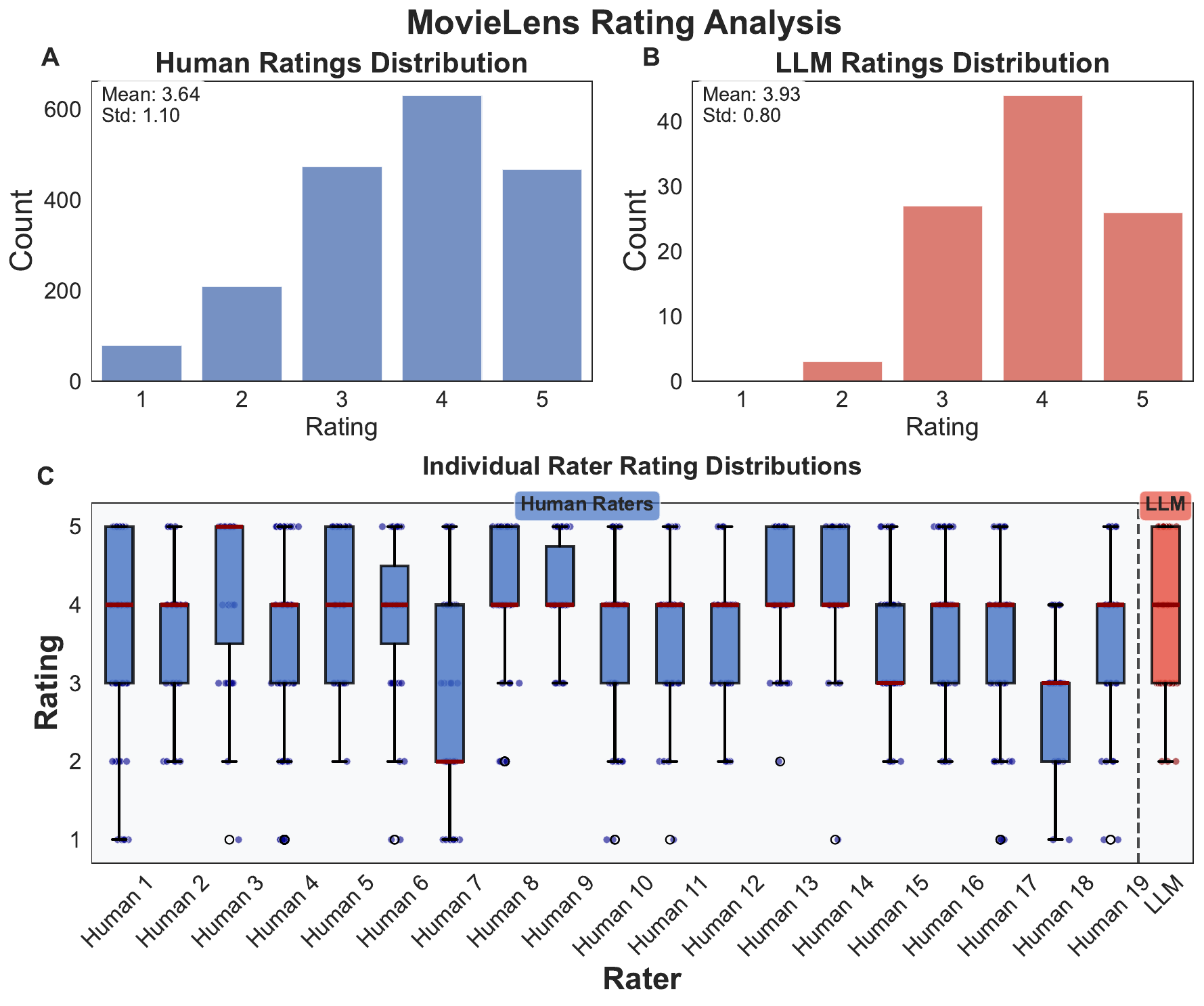}
    \caption{MovieLens ratings distribution for human and LLM.
    }
    \label{fig:movielens_rating_distribution}
\end{figure}

\myparagraph{Equivalence Testing Outcomes.}
We assessed whether LLM-substituted annotations were statistically equivalent to human annotations across varying bootstrap sample sizes \(B\). As shown in ~\autoref{fig:movielen_p_evlauation} and ~\autoref{fig:Politifact_p_evaluation}, the p-values exhibit different behaviors across datasets: on \textbf{MovieLens 100K}, p-values tend to decrease with larger \(B\), indicating increased stability; on \textbf{PolitiFact}, no consistent pattern emerges.

Table~\ref{tab:p value result} summarizes the final results at \(B = 300\), reporting the mean and standard deviation across 10 runs for the equivalence margin, Krippendorff’s $\alpha$ for the human group, the LLM-substituted group, and the randomly substituted group, along with the corresponding TOST p-values ($p_1$ and $p_2$) for both LLM and random substitutions. The LLM passed the equivalence test on the MovieLens 100K dataset but failed on PolitiFact—despite nearly identical $\alpha$ scores between the LLM- and human-only groups. This highlights how small margins and higher variability, shown in the per-rater results discussed subsequently, can lead to non-equivalence conclusions. In contrast, substituting annotations with random labels consistently produced significantly lower $\alpha$ scores and failed the equivalence test in both datasets, confirming that LLM-generated annotations are substantially more aligned with human judgment than random labels.

\myparagraph{Annotation Distribution Comparisons.}
 and Figure~\ref{fig:movielens_rating_distribution} show the rating distributions across annotators. LLM ratings align more closely with human annotations in MovieLens 100K, while notable differences are observed in PolitiFact. These distributional patterns mirror the statistical equivalence findings.

\myparagraph{Agreement Change per Rater.}
Finally, we examined how Krippendorff’s $\alpha$ changed when each individual human coder was replaced with the LLM (Figure~\ref{fig:change_alpha_movielen},~\ref{fig:politifact_alpha_change}). For the PolitiFact dataset, changes ranged from –2.3\% to 2.5\%, and for MovieLens 100K, from –1.5\% to 1.7\%. In both datasets, some raters showed minimal change (as low as 0.1\%), suggesting that the LLM closely aligned with certain individuals. The variability in the per rater change in Krippendorff’s $\alpha$ suggests that while substitution effects are minimal on average, individual rater alignment with the LLM may vary. Additionally, the smaller variability in the changes of the per rater Krippendorf $\alpha$ on the MovieLens 100k dataset dataset compared to the PolitiFact dataset also indicates a closer LLM alignment to the raters for the movie rating task compared to rating the truthfulness of a piece of information.

\begin{figure}[t]
    \centering
        \includegraphics[width=\linewidth]{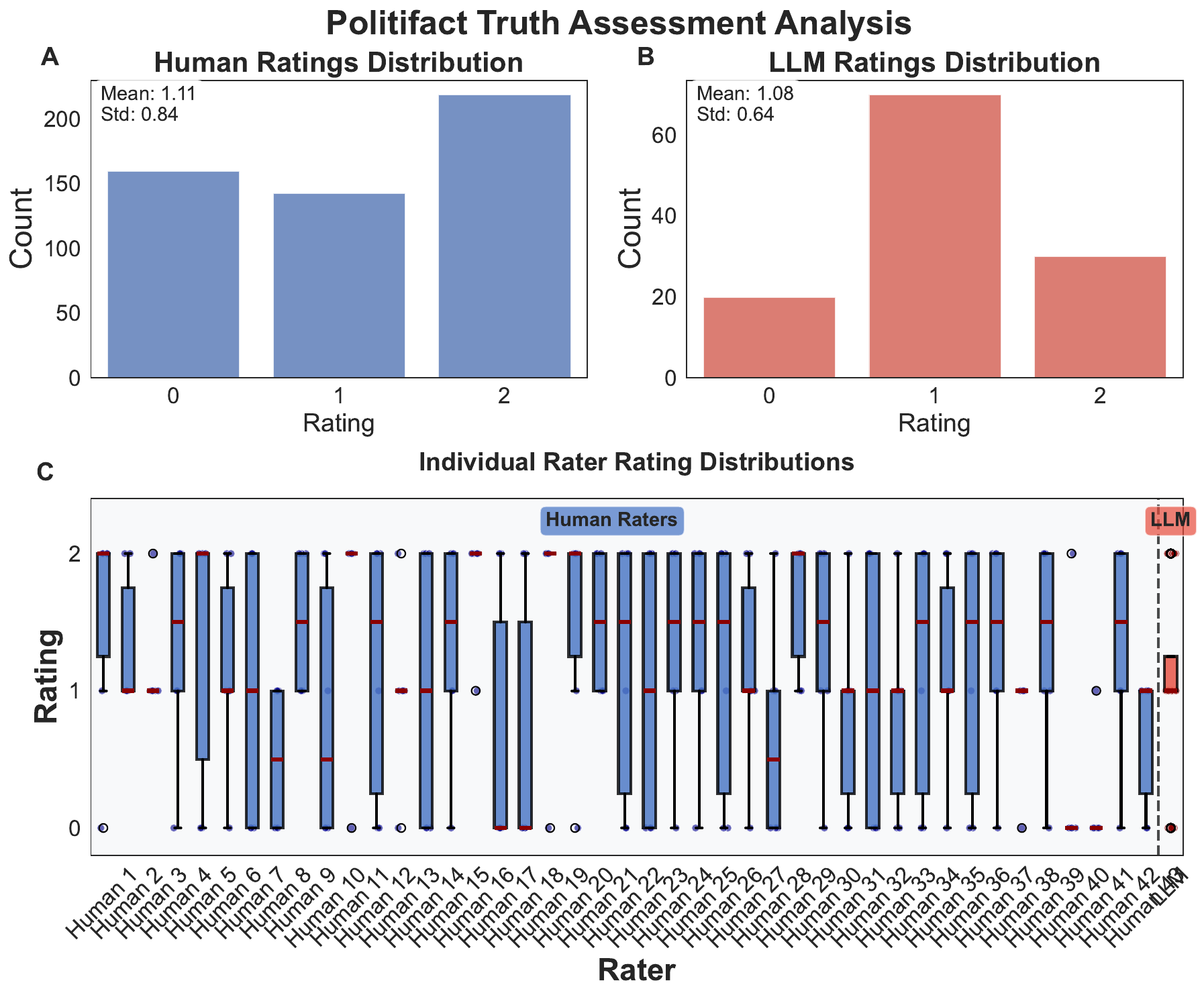}
    \caption{PolitiFact ratings distribution for human and
LLM.}
    \label{fig:politifact_rating_distribution}
\end{figure}

\section{Discussion}

Our results demonstrate that the LLM’s ability to substitute for human annotators is highly task-dependent. On the MovieLens dataset, the model passed our equivalence test, whereas on the PolitiFact dataset it failed. This divergence reflects fundamental differences between the two annotation domains: preference-based rating versus fact-checking.

\begin{figure}[t]
    \centering
        \includegraphics[width=\linewidth]{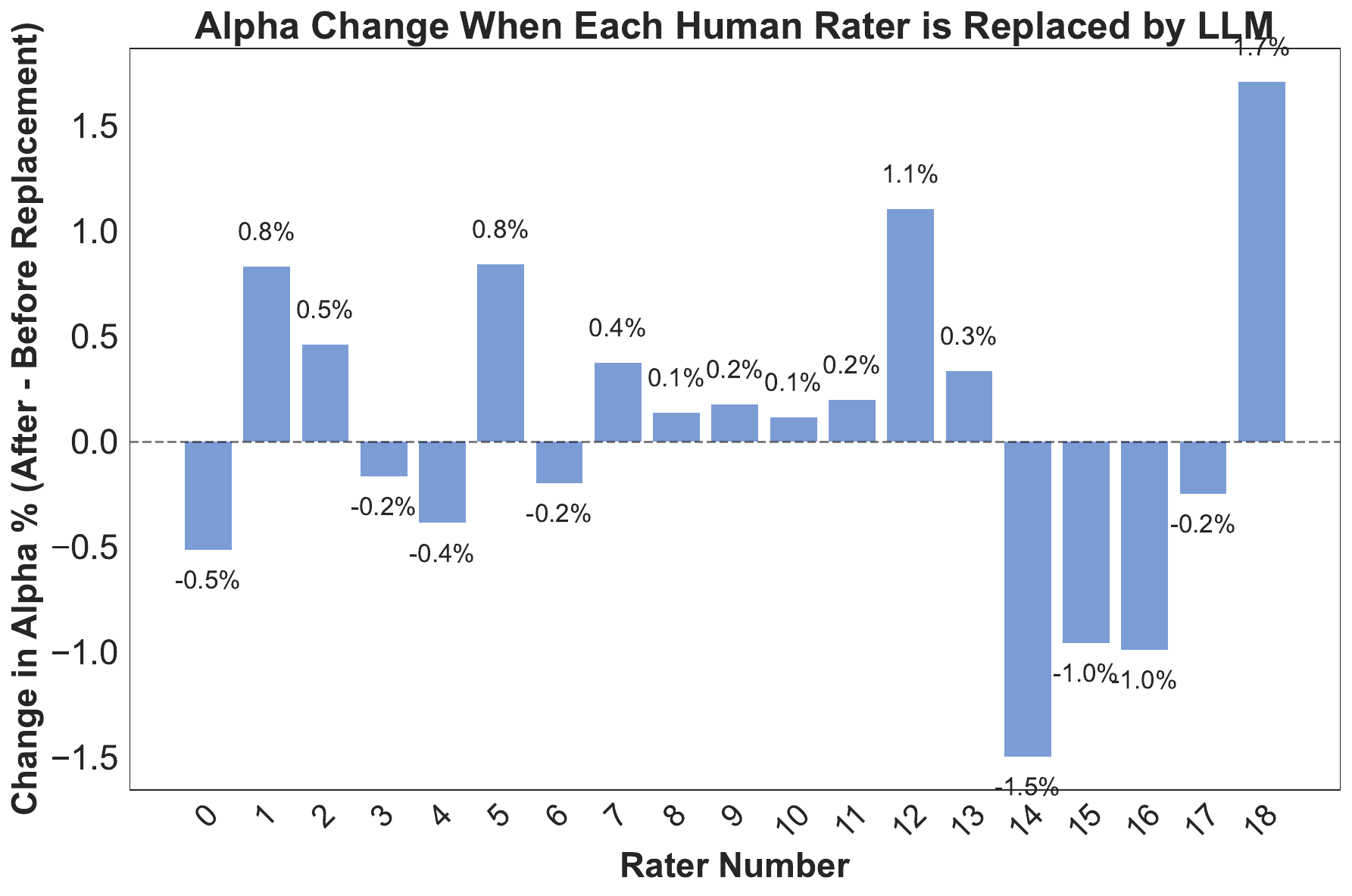}
    \caption{The change in Krippendorf's $\alpha$ before and after replacing for Movielens 100K dataset.
    }
    \label{fig:change_alpha_movielen}
\end{figure}

For MovieLens, it’s not surprising that LLM ratings align with human ratings. As shown in~\autoref{fig:movielens_rating_distribution}, the model’s scores follow the general patterns of human raters. This likely comes from its pretraining on large amounts of movie-related text~\cite{di2025llms}, which helps it learn common rating habits and genre cues. In other words, judging movie enjoyment mostly means recognizing shared cultural opinions and repeating them. While there is some subjectivity, the range of reasonable answers is limited and well covered in the training data. That’s why replacing human ratings with LLM ratings hardly changes Krippendorff’s $\alpha$ in~\autoref{fig:change_alpha_movielen}.

\begin{figure}[t]
    \centering
        \includegraphics[width=\linewidth]{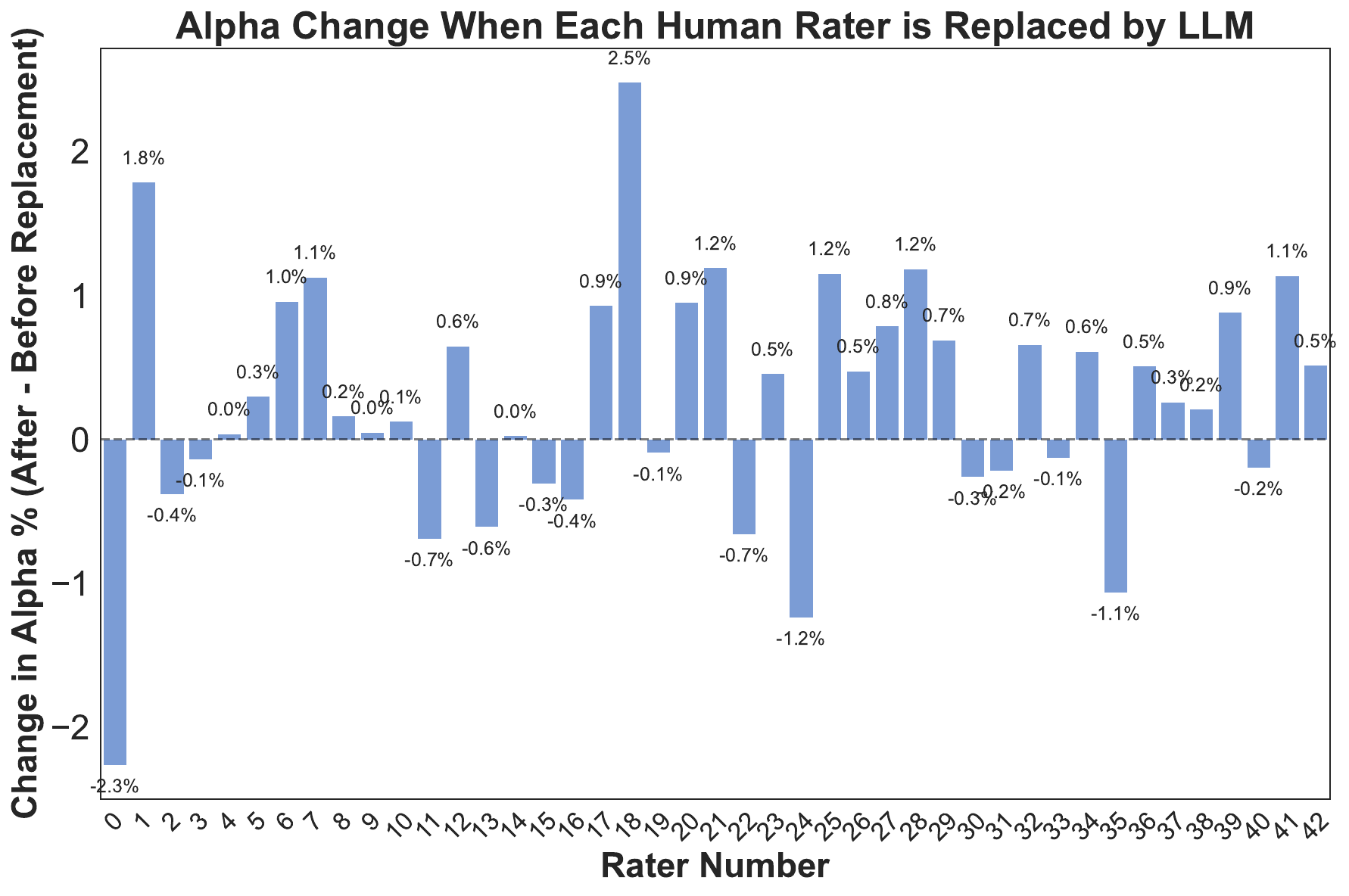}
    \caption{The change in Krippendorf's $\alpha$ before and after replacing for PolitiFact dataset. 
    }
    \label{fig:politifact_alpha_change}
\end{figure}

In the PolitiFact dataset, replacing human ratings with LLM ratings lowers reliability across raters (see~\autoref{fig:politifact_alpha_change}). Fact-checking is harder than movie rating because it requires domain knowledge, evidence usage, and political framing awareness. Human annotators bring in their own beliefs, expertise, and even mood, which creates variability~\cite{roitero2020can, jensen2016international, gilbert1993you}. Different cues like familiarity, political views, or emotional reactions influence people in different ways~\cite{hammond1955probabilistic, hoffman1960paramorphic}. The LLM, however, takes a narrower and cautious approach: it avoids strong labels like “True” or “False” and stays near the middle categories~\autoref{fig:politifact_rating_distribution}. This mismatch with human judgment patterns explains why inter-rater reliability drops more in this case.

The alpha-change analysis (Figures~\ref{fig:change_alpha_movielen} and~\ref{fig:politifact_alpha_change}) shows a clear contrast. In MovieLens, swapping some human raters with the LLM has a minimal impact on reliability, indicating that the model can effectively mimic certain annotators. In PolitiFact, though, replacements cause big drops in agreement for some raters. This suggests LLMs still have trouble replicating the unique, knowledge-based reasoning humans use in complex or disputed topics.

Taken together, these findings underscore two key points. First, our evaluation method captures meaningful differences across tasks: it is mathematically rigorous yet sensitive to domain characteristics. Second, LLM substitution is more viable for preference-oriented annotations than for knowledge-intensive or adversarial tasks. These results highlight the importance of aligning LLM-based annotation strategies with the epistemic demands of the domain.

\myparagraph{Guidelines}

Practitioners should apply our evaluation methodology as follows. For existing datasets, replicate the workflow described in Section~\ref{sec:experimental methodology}. If the dataset does not meet our minimum criteria (four annotators, at least 80 items), a small-scale annotation effort should be undertaken. The same approach applies to new tasks: collect modest and diverse annotations, compute reliability measures, and test LLM substitution feasibility before scaling.

\myparagraph{Interpreting the Results}

When an LLM passes the equivalence test, it can be considered a candidate substitute for the replaced annotator. If that annotator is a gold-standard rater, the LLM may then be deployed to expand annotations at scale. If substitution fails, human annotation remains indispensable, though alternative models or prompting strategies could be tested. Importantly, our alpha-change framework also enables more granular exploration: identifying which human rater is most closely approximated, comparing across LLMs, or diagnosing systematic biases in annotation behavior. This level of analysis extends beyond a simple pass/fail judgment, offering a roadmap for responsibly integrating LLMs into annotation pipelines.

\section{Conclusion} 
Our experimental results demonstrate that the proposed evaluation method can effectively assess whether an LLM approximates human judgment in specific text annotation tasks, using only a small number of annotated items. We applied the method to two datasets—MovieLens 100K ($p = 0.004$) and PolitiFact ($p = 0.155$)—with differing outcomes. The LLM passed the equivalence test on MovieLens 100K but not on PolitiFact. These results highlight that performance is not consistent across tasks and depends heavily on the nature of the annotation task. 

This method provides a practical way to detect differences in annotation behavior between humans and LLMs. It also offers an opportunity to evaluate a small set of annotations before deciding whether to use an LLM for large-scale annotations.

\myparagraph{Limitations and Future Work} 

The performance of the LLM is likely to vary depending on the specific model and the prompts used. Future work should investigate how various LLM architectures and prompt strategies impact the results, facilitating a more comprehensive evaluation of LLM-human alignment across diverse annotation contexts.

\begin{acks}
This research was conducted by the ARC Centre of Excellence for Automated Decision-Making and Society (ADM+S, \grantnum{ARC}{CE200100005}), and funded fully by the Australian Government through the \grantsponsor{ARC}{Australian Research Council}{https://www.arc.gov.au/}.
 This work was conducted on the unceded lands of the  Woi wurrung and Boon wurrung language groups of the eastern Kulin Nation. We pay our respect to Ancestors and Elders, past and present, and extend that respect to all Aboriginal and Torres Strait Islander peoples today and their connections to land, sea, sky, and community. 
\end{acks}

\bibliographystyle{ACM-Reference-Format}
\bibliography{bibs/IR}

\end{document}